
\input phyzzx
\titlepage

\title{
Dual Ginzburg-Landau Theory for Quark Confinement
and Dynamical Chiral-Symmetry Breaking
\foot{
To apper in Proc. of the International Symposium on
``Intermediate Energy Nuclear Physics'',
Beijing China, August 1994.
}
}

\author{
H.~Toki$^{\rm a,b,c}$, H.~Suganuma$^{\rm a}$, S.~Sasaki$^{\rm b}$
and H.~Ichie$^{\rm c}$}

\address{ {\rm a)}
The Institute of Physical and Chemical Research (RIKEN), \break
Wako, Saitama 351-01, Japan }

\address{ {\rm b)}
Research Center for Nuclear Physics (RCNP), \break
Osaka University, Ibaraki, Osaka 567, Japan }

\address{ {\rm c)}
Department of Physics, Tokyo Metropolitan University, \break
Hachiohji, Tokyo 192, Japan }

\abstract{
Nonperturbative features of QCD are studied using the dual
Ginzburg-Landau (DGL) theory with QCD-monopoles.
The linear quark potential appears in the QCD-monopole
condensed vacuum.
We find that QCD-monopole condensation plays an essential role to
the dynamical chiral-symmetry breaking.
We also investigate the QCD phase transition at finite temperature
in the DGL theory.
}

\endpage

\sequentialequations

\noindent
{\bf 1. Introduction}

\vskip 0.3cm

The asymptotic freedom of QCD enables us to use
the perturbative QCD calculation in the ultraviolet region
as the deep inelastic scattering,
but it leads to a strong coupling system
in the infrared region of the hadron physics.
This strong interaction provides color confinement and
also dynamical chiral-symmetry breaking (D$\chi $SB)
as the nonperturbative features of QCD.
In particular, color confinement is one of the most unique
features of the nonperturbative QCD,
and therefore the understanding of the confinement mechanism is
the central issue for the hadron physics.
The D$\chi $SB is also an important feature in the nonperturbative QCD.
Although recent lattice QCD studies support a close relation
between color confinement and D$\chi $SB, no clear physical
interpretation has been presented yet.
In this paper,
we investigate these nonperturbative properties of QCD
in terms of the dual Ginzburg-Landau theory
\REF\suganuma{
H.~Suganuma, S.~Sasaki and H.~Toki,
Nucl.~Phys.~{\bf B435} (1995) 207. \nextline
H.~Toki, H.~Suganuma and S.~Sasaki,
Nucl.~Phys.~{\bf A577} (1994) 353c.
}
\REF\suzuki{
T.~Suzuki, Prog.~Theor.~Phys. {\bf 80} (1988) 929 ;
{\bf 81} (1989) 752. \nextline
S.~Maedan and T.~Suzuki, Prog.~Theor.~Phys. {\bf 81} (1989) 229.
}
[\suganuma,\suzuki],
an infrared effective gauge theory of QCD based
on the dual Higgs mechanism.

\vskip 0.5cm

\noindent
{\bf 2. Dual Ginzburg-Landau Theory}

\vskip 0.3cm

In 1981, 't~Hooft presented an interesting fact
that a non-abelian gauge theory reduces to an abelian
gauge theory with color-magnetic monopoles by the abelian gauge fixing
\REF\thooft{
G.~'t~Hooft, Nucl.~Phys.~{\bf B190} (1981) 455.
}[\thooft].
The appearance of these color-magnetic monopoles in QCD, called as
QCD-monopoles, would be important for the confinement mechanism,
because the quark confinement can be interpreted by
condensation of such monopoles as shown below.
Since QCD-monopoles behave as singularities of vector potential $A_\mu $,
we cannot introduce $A_\mu$ in the standard way.
Instead, we can formulate the gauge theory without
singularity following Zwanziger
\REF\zwanziger{
D.~Zwanziger, Phys.~Rev.~{\bf D3} (1971) 880.
}[\zwanziger]
by introducing the dual gauge field
$B_\mu$, which satisfies
$\partial_\mu  B_\nu -\partial_\nu  B_\mu =\ ^* F_{\mu \nu }$.
In this formalism, the duality of the gauge theory becomes manifest.
The QCD-monopole current $k_\mu$ couples
with the dual gauge field as $k_\mu B^\mu$
in the similar way as the ordinary color-electric current,
$j_\mu A^\mu$.
The self-interaction among QCD-monopoles is
introduced to realize QCD-monopole condensation,
which is strongly supported by the recent studies based on the
lattice QCD
\REF\shiba{
A.~S.~Kronfeld, G.~Shierholz and U.~-J.~Wiese,
Nucl.~Phys.{\bf B293} (1987) 461. \nextline
T.~Suzuki and I.~Yotsuyanagi, Phys.~Rev.{\bf D42} (1990) 4257.
}
[\shiba].

The Lagrangian of the dual Ginzburg-Landau theory (DGL) is
given as [\suganuma,\suzuki]
$$
\eqalign{
{\cal L}_{\rm DGL}=
&-{1 \over 2n^2}
[n\cdot (\partial \wedge \vec A)]^\nu
[n\cdot ^*(\partial \wedge \vec B)]_\nu
+{1 \over 2n^2}
[n\cdot (\partial \wedge \vec B)]^\nu
[n\cdot ^*(\partial \wedge \vec A)]_\nu \cr
&-{1 \over 2n^2}
[n\cdot (\partial \wedge \vec A)]^2
-{1 \over 2n^2}
[n\cdot (\partial \wedge \vec B)]^2
+\bar q (i\partial \kern -2mm / -e \vec A \kern -2mm / \cdot \vec H-m)q \cr
&+\sum_{\alpha =1}^3[|(i\partial_\mu -g \vec \epsilon _\alpha
\cdot \vec B_\mu )\chi _\alpha |^2
-\lambda (|\chi _\alpha |^2-v^2)^2]
}
\eqn\DGLlag
$$
with
$\vec A^\mu =(A^\mu _3, A^\mu _8)$,
$\vec B^\mu =(B^\mu _3, B^\mu _8)$ and $\vec H=(T_3,T_8)$.
Here, $g$ is the unit magnetic charge obeying the Dirac condition
$eg=4\pi $, and $\vec \epsilon_\alpha $ denotes the relative
magnetic charge of the QCD-monopole field $\chi _\alpha $ ($\alpha $=1,2,3).
The dual Meissner effect is caused by QCD-monopole condensation due to
the self-interaction of $\chi _\alpha $ in the case $v^2>0$.
It provides the mass of the dual gauge field $ B_\mu$.
In the QCD-monopole condensed vacuum, the DGL Lagrangian becomes
$$
\eqalign{
{\cal L}_{\rm MF}
=-{1 \over 2n^2}
[n\cdot (\partial &\wedge \vec A)]^\nu
[n\cdot ^*(\partial \wedge \vec B)]_\nu
+{1 \over 2n^2}
[n\cdot (\partial \wedge \vec B)]^\nu
[n\cdot ^*(\partial \wedge \vec A)]_\nu \cr
-{1 \over 2n^2}
[n\cdot (\partial \wedge \vec A)]^2&
-{1 \over 2n^2}
[n\cdot (\partial \wedge \vec B)]^2
+\bar q (i\partial \kern -2mm / -e \vec A \kern -2mm / \cdot \vec H-m)q
+{1 \over 2}m_B^2 \vec B^2
}
\eqn\DGLMF
$$
with $m_B= \sqrt 3 gv$.
The QCD-monopole also becomes massive as $m_\chi  = 2\sqrt\lambda v$.
Similar to the Higgs mechanism in the superconductivity,
the color-electric field is then excluded in the QCD vacuum
through the dual Meissner effect, and is
squeezed between color sources to form the hadron flux tube.

\vskip 0.5cm

\noindent
{\bf 3. Quark Confinement Potential}

\vskip 0.3cm

We investigate the quark confinement in terms of the
linear inter-quark potential, which is supported by the
lattice QCD in the quenched approximation.
By integrating over $A_\mu$ and $ B_\mu$ in the partition
functional of the DGL theory, the current-current correlation
is obtained as
$$
{\cal L}_j=-{1 \over 2}\vec j_\mu D^{\mu \nu }\vec j_\nu
\eqn\JJ
$$
with the nonperturbative gluon propagator,
$$
D_{\mu \nu }={1 \over \partial^2}\{g_{\mu \nu }+(\alpha _e-1)
{\partial_\mu \partial_\nu \over \partial^2} \}
-{1 \over \partial^2}
{m_B^2 \over \partial^2+m_B^2}
{1 \over (n\cdot \partial)^2}
\epsilon ^\lambda  \ _{\mu \alpha \beta }\epsilon _{\lambda \nu \gamma \delta
}n^\alpha n^\gamma \partial^\beta \partial^\delta
\eqn\GLa
$$
in the Lorentz gauge.
Putting a static quark with color charge $\vec Q$ at ${\bf x}={\bf a}$
and a static antiquark with color charge
$-\vec Q$ at ${\bf x}={\bf b}$, the quark current is written as
$
\vec j_\mu (x)=\vec Qg_{\mu 0}\{\delta ^3({\bf x}-{\bf b})
-\delta ^3({\bf x}-{\bf a})\}.
$
We finally obtain the inter-quark potential including the Yukawa and
linear terms,
$$
V(r)=-{\vec Q^2 \over 4\pi }{e^{-m_Br}\over r}
+{\vec Q^2 m_B^2 \over 8\pi }\ln({m_B^2+m_\chi ^2 \over m_B^2})\cdot r,
\eqn\POT
$$
where $r=|{\bf r}| =|{\bf b}-{\bf a}|$ is the relative distance.
Here, we have identified ${\bf n} // {\bf r}$,
which is also used in the similar context of the dual string theory
\REF\nambu{
Y.~Nambu, Phys.~Rev.~{\bf D10} (1974) 4262.
}[\nambu], because of
the axial symmetry of the system
and the energy minimum condition. Otherwise,
the energy of the system diverges.
It should be noted that
the expression for the string tension,
the coefficient of the linear potential, agrees with the one for
the energy per length of the vortex
in the type-II superconductor.

We compare the static potential with the
phenomenological one, for example,
the Cornell potential.
We get a good agreement as shown in Fig.1
with the choice of $e=5.5$, $m_B=0.5{\rm GeV}$
and $m_\chi =1.26{\rm GeV}$ corresponding to
$\lambda=25$ and $v=126{\rm MeV}$,
which provide $k$=1.0GeV/fm for the
string tension and the radius of the hadron flux as
$m^{-1}_B=0.4{\rm fm}$.

\vskip 0.5cm

\noindent
{\bf 4. Dynamical Chiral-Symmetry Breaking}

\vskip 0.3cm

The dynamical chiral-symmetry breaking (D$\chi $SB) is also important
for the hadron properties as well as color confinement.
We discuss here the D$\chi $SB in terms of the mass generation
of light quarks in the QCD-monopole condensed vacuum
\REF\sasaki{
S.~Sasaki, H.~Suganuma and H.~Toki,
preprint, RIKEN-AF-NP-172 (1994).
}
[\suganuma,\sasaki].
We formulate the Schwinger-Dyson (SD) equation for massless quark as
$$
S^{-1}_q(p)={p \kern -2mm /}+ \int {d^4k \over i(2\pi )^4}
\vec Q^2 \gamma ^\mu S_q(k)\gamma ^\nu
D_{\mu \nu }^{\rm sc}(k-p),
\eqn\SDEa
$$
where we assume the quark propagator $S_q(p)$ as
$
S_q(p)^{-1}={p \kern -2mm /} -M(-p^2)+i\eta .
$
In the presence of light quarks,
there appears the screening effect
in the long distance of the linear potential corresponding to
the cut of the hadron flux tube
by the light-quark pair creation.
Hence, we introduce the corresponding infrared cutoff
$a$ in the gluon propagator $D_{\mu \nu }^{\rm sc}$ by
the replacement,
${1 \over (n \cdot k)^2}
\rightarrow {1 \over (n \cdot k)^2 +a^2}$ [\suganuma].

Taking the trace and making the Wick rotation in the $k_0$-plane,
we obtain the SD equation in the Euclidean metric,
$$
M(p^2)= \int{d^4k \over (2\pi )^4} \vec Q^2 {M(k^2) \over k^2+M^2(k^2)}
D_\mu ^{\mu {\rm sc}}(k-p),
\eqn\SDEb
$$
where the gluon propagator including the screening effect is given as
$$
D_\mu ^{\mu     {\rm sc}}(k) =
        {1 \over (n \cdot k)^2+a^2} \cdot {1 \over k^2} \cdot
        {2m_B^2 \over k^2+m_B^2}\{k^2-(n \cdot k)^2 \}
        +{3+\alpha _e \over k^2}
\eqn\GP
$$
in the Lorentz gauge.
After performing the angular integral,
we obtain the final expression for the SD equation,
$$
\eqalign{
M(p^2)
 &=\int_0^\infty {dk^2 \over 16\pi ^2} {\vec Q^2 M(k^2)
 \over k^2+M^2(k^2)} \biggl(
 {4k^2  \over   k^2+p^2+m_B^2 + \sqrt{(k^2+p^2+m_B^2)^2-4k^2p^2} }
 \cr
 &+{(1+\alpha _e)k^2 \over {\rm max}(k^2, p^2)}
 +{1 \over \pi p_{_T}}
 \int_{-k}^k dk_n {1 \over \tilde k_n^2+a^2} \cr
 &\times [
 (m_B^2-a^2) \ln \{ { {\tilde k_n^2+(k_{_T}+p_{_T})^2+m_B^2
 \over \tilde k_n^2+(k_{_T}-p_{_T})^2+m_B^2} } \}
 +a^2   \ln \{ { {\tilde k_n^2+(k_{_T}+p_{_T})^2
 \over \tilde k_n^2+(k_{_T}-p_{_T})^2} } \}     ] \biggr)
}
\eqn\SDEc
$$
with
$\tilde k_n \equiv k_n-p_n$ and  $k_{_T} \equiv (k^2-k_n^2)^{1/2}$.

In solving the SD equation, we use the Higashijima-Miransky
ansatz
\REF\higashijima{
K.~Higashijima, Phys.~Rev.~{\bf D29} (1984) 1228;
Prog.~Theor.~Phys. Suppl. {\bf 104} (1991) 1. \nextline
V.~A.~Miransky, Sov.~J~.~Nucl.~Phys.~{\bf 38}(2) (1983) 280.
}
with a hybrid type of the running coupling constant,
$$
\tilde e=e({\rm max}\{p^2, k^2\}), \quad
e^2(p^2)=
{48\pi ^2 (N_c+1) \over (11N_c-2N_f)
\ln\{(p^2+p^2_c)/\Lambda ^2_{\rm QCD}\}}.
\eqn\HMA
$$
Here, $p_c$ is defined as
$p_c\equiv \Lambda _{\rm QCD}\exp[{24 \pi ^2 \over e^2}\cdot
{N_c+1 \over 11N_c-2N_f}]$ with $e = e(0)$.
This ansatz naturally connects to the asymptotic freedom of the
running coupling at large momentum.
The coupling constant at low energy, $e(p^2\sim 0) \simeq e$,
controls the strength of the linear confinement potential.

Fig.2 shows the quark mass function $M(p^2)$ with
$e$=5.5 and $a =80{\rm MeV}$.
The QCD scale parameter is set to a realistic value
$\Lambda _{\rm QCD}=200{\rm MeV}$.
In order to see the effect of QCD-monopole condensation,
we vary the mass of the dual gauge field, $ m_B $.
There is no non-trivial solution for the case with small
$m_B < 300{\rm MeV}$.
A non-trivial solution is barely obtained at $ m_B=300{\rm MeV}$,
and $M(p^2)$ increases rapidly with $ m_B $ as shown in Fig.2.
Hence, QCD-monopole condensation provides a crucial
contribution to D$\chi $SB.

Taking the value for the mass of the dual gauge field as
$ m_B=0.5{\rm GeV}$ extracted from the
linear potential, we get the result for $M(p^2)$ as shown in Fig.3.
The quark mass function $M(p^2)$ in the space-like region
is directly obtained from the SD equation.
We extrapolate $M(p^2)$ into the time-like region
using a polynomial function as a simulation
of the analytic continuation.
This curve does not cross
the on-shell condition $M^2(p^2)+p^2=0$ ($p_\mu $: Euclidean momentum)
and hence the quark propagator does not have a physical pole.
This may indicate the light-quark confinement.

We calculate the several quantities related to D$\chi $SB
from the solution of the SD equation.
The constituent quark mass in the infrared region
is found to be $M(0)$=348MeV.
The quark condensate is obtained as
$\langle \bar {q}q \rangle =-(229{\rm MeV})^3$.
The pion decay constant is also calculated as
$f_\pi $=83.6MeV using the Pagels-Stoker formula
\REF\ps{
H.~Pagels and S.~Stoker, Phys.~Rev.~{\bf D20} (1979) 2947;
{\bf D22} (1980) 2876.
}[\ps].
These values are to be compared with the standard values;
$M$(0)=350 MeV, $\langle \bar{q}q \rangle
=-(225 \pm{\rm50} {\rm MeV})^3$ and $f_{\pi}=93{\rm MeV}$.

\vskip 0.5cm

\noindent
{\bf 5. QCD Phase Transition at Finite Temperature}

\vskip 0.3cm

The DGL theory is now able to describe many interesting quantities.
Here, we study the change of the QCD vacuum at finite temperature
\REF\ichie{
H.~Ichie, H.~Suganuma and H.~Toki,
preprint, RIKEN-AF-NP-176 (1994).
}
[\ichie] in terms of QCD-monopole condensation.
To concentrate on the confinement properties, we consider the pure
gauge case, where the quark degrees of freedom are frozen.
In this case, we can drop the quark term in the DGL Lagrangian
and perform integration over the gauge field $A_\mu $.
Hence, we obtain the partition functional as
$$
  Z[J] = \int {\cal D}{\chi_{\alpha}}{\cal D}{\vec{B}_{\mu}}
\exp{\left( i\int d^4x \{{\cal{L}}_{\rm DGL}
-J\sum_{\alpha=1}^3|\chi_\alpha|^2\} \right) },
\eqn\Zj
$$
where ${\cal L}_{\rm DGL}$ has a simple form,
$$
 {\cal L}_{\rm DGL}= - {1 \over 4}
(\partial_{\mu}\vec{B}_{\nu}-\partial_{\nu}\vec{B}_{\mu})^2 +
 \sum_{\alpha=1}^3[|(i\partial_{\mu}-g\vec{\epsilon_{\alpha}}
\cdot\vec{B}_{\mu})
\chi_{\alpha}|^2 - \lambda(|\chi_{\alpha}|^2-v^2)^2].
\eqn\La
$$
Here, we have introduced the quadratic source term instead of the
linear source term, which is commonly used.
Such an introduction of the quadratic source term is quite powerful
for the formulation of the effective potential, especially
in the negative-curvature region of the classical potential,
where the use of the linear source term does not work well.

The effective potential at finite temperature,
which physically corresponds to the thermodynamical potential,
is then obtained as
$$
\eqalign{
V_{\rm eff}(\bar \chi ;T) =   3 \lambda ( \bar \chi^2 - v^2 )^2
          &+ 3 {T \over \pi^2} \int_0^\infty  dk k^2 \ln{
           \left(  1 - e^{ - \sqrt{ k^2 + m_B^2}/T }
           \right)  } \cr
          &+ {3 \over 2} {T \over \pi^2}\int_0^\infty  dk k^2 \ln{
           \left(  1 - e^{ - \sqrt{ k^2 + m_{\chi}^2}/T }
           \right)  }.
}
\eqn\Vb
$$
Here, the masses of the QCD-monopole and the dual gauge field
depend on the QCD-monopole condensate $\bar \chi $,
$$
   m_\chi ^2(\bar \chi ) = 2\lambda (3 \bar \chi ^2-v^2) + J(\bar \chi ) =
4\lambda \bar \chi ^2,
   \hbox{\quad} m_B^2(\bar \chi ) = 3 g^2 \bar \chi ^2.
\eqn\Ma
$$

We provide the calculated results on the effective potential
in Fig.4.
At $T=0$, one minimum appears at a finite $\bar \chi $,
which corresponds to the condensed phase of QCD-monopoles.
As the temperature increases, the minimum moves toward a small
$\bar \chi $ value, and the second minimum appears at $\bar \chi =0$
above the lower critical temperature $T_{\rm low} \simeq 0.39$GeV.
The potential values at the two minima become equal at
$T_c \simeq 0.49$GeV, which corresponds to the thermodynamical
critical temperature for the
QCD phase transition.
In this case, it is of first order.
Then the trivial vacuum stays as the absolute minimum above the
critical temperature.
When light dynamical quarks are included,
we also expect the chiral-symmetry restoration as well as the
deconfinement phase transition at this critical temperature,
because QCD-monopole condensation is essential
for D$\chi $SB as demonstrated in the previous sections.

Since the critical temperature seems too high in the above discussion,
we consider the temperature dependence of
the monopole coupling constant $\lambda $.
This is very probable because the interaction among QCD-monopoles
is weakened at finite temperature due to the asymptotic
free behavior of QCD.
Hence, we examine a simple case where $\lambda $ decreases linearly with $T$,
$$
  \lambda (T) = \lambda  \left( {T_c - \alpha T \over T_c}  \right),
\eqn\Lam
$$
where a constant $\alpha =0.96$ is chosen so as to satisfy $T_c=0.2$GeV.
We find in this case also a weak first order phase transition.
Here, we are able to compare with the numerical
results with the pure-gauge lattice QCD on the string tension
$k(T)$
\REF\gao{ M.~Gao, Nucl.~Phys.~{\bf B9} (Proc. Suppl.) (1988) 368.}
[\gao].
The lattice QCD results are shown by black dots below the
critical temperature in Fig.5, while the results for the variable
$\lambda (T)$ go through the dots.
We also find that the masses of glueballs (QCD-monopoles,
the dual gauge fields) drop largely toward
$T_c$ with $T$ from those of order of 1 GeV at zero temperature.
It would be very interesting to check this phenomena
by the lattice QCD and also by experiment.

\vskip 0.5cm

\noindent
{\bf 6. Summary}

\vskip 0.3cm

We have studied the dual Ginzburg-Landau (DGL) theory as the effective
theory of QCD.
QCD is reduced to an abelian gauge theory with QCD-monopoles in
't~Hooft's abelian gauge.
According to QCD-monopoles condensation,
the dual Higgs mechanism works as the mass generation of the
dual gauge field.
We have derived the static inter-quark potential in the DGL theory.
In the QCD-monopole condensed vacuum, there appears
the linear potential responsible for the quark confinement.

We have also studied the dynamical chiral-symmetry breaking (D$\chi $SB)
in the DGL theory.
We find an essential role of QCD-monopole condensation to D$\chi $SB.
With the parameters extracted from the quark confining potential,
we obtain reasonable values
for the constituent quark mass $M(p=0)$=348MeV, the quark condensate
$\langle \bar{q}q \rangle =-(229{\rm MeV})^3$ and
the pion decay constant $f_\pi $=83.6MeV.

The DGL theory predicts the existence of an axial-vector particle
$B_\mu $ and the scalar QCD-monopole $\chi _\alpha $ with masses of
$ m_B \sim 0.5{\rm GeV}$ and
$ m_\chi  \sim 1.5{\rm GeV}$ with admittedly a
large error of about 1GeV. It would be important to look for these
particles in the hadron spectra. Theoretically, we are investigating
the decay properties of these particles.

We have discussed also the properties of the QCD vacuum
at finite temperature in terms of the DGL theory.
We find a first order deconfinement phase transition in the
pure gauge case. We are now making an effort to introduce
dynamical quarks in the discussion of the QCD phase transition.

\vskip1cm

\refout

\endpage

\centerline{\fourteenpoint Figure Captions}

\vskip1cm

\item{Fig.1.}
The static quark potential $V(r)$ in the
dual Ginzburg-Landau theory.
The dashed curve denotes the Cornell potential.

\item{Fig.2.}
The dynamical quark mass $M(p^2)$
as a function of the
Euclidean momentum squared $p^2$ for $m_B$=300, 400 and 500 MeV.

\item{Fig.3.}
The dynamical quark mass squared $M^2(p^2)$
as a function of $p^2$.
The dotted straight line denotes the on-shell state.

\item{Fig.4.}
The effective potentials at various temperatures
as functions of the QCD-monopole condensate $\bar \chi $.
The crosses denote their minima.

\item{Fig.5.}
The string tensions $k(T)$ as functions of the
temperature $T$ for a constant $\lambda$ and a variable $\lambda (T)$.
The lattice QCD results in the pure gauge
are shown by black dots.

\end